\documentclass[american,aps,pra,reprint,tightenlines,superscriptaddress,twocolumn,twoside,longbibliography]{revtex4-1}
\pdfoutput=1

\usepackage[utf8]{inputenc}

\usepackage[unicode=true,pdfusetitle, bookmarks=true,bookmarksnumbered=false,bookmarksopen=false, breaklinks=false,pdfborder={0 0 0},backref=false,colorlinks=false] {hyperref}
\hypersetup{ colorlinks,linkcolor=myurlcolor,citecolor=myurlcolor,urlcolor=myurlcolor}
\usepackage{braket,colortbl,amsthm,amsmath,amssymb,dsfont}
\definecolor{myurlcolor}{rgb}{0,0,0.7}
\usepackage{graphics,graphicx,relsize}
\usepackage{color}
\usepackage{XCharter}
\usepackage[T1]{fontenc}
\usepackage{mathptmx}
\usepackage[left=1.6cm,right=1.6cm,top=2.45cm,bottom=2.45cm]{geometry}
\usepackage[caption=false]{subfig}
\usepackage{bigints}
\usepackage{accents}
 
\theoremstyle{plain}

\def\bea{\begin{eqnarray}}
\def\eea{\end{eqnarray}}
\def\ba{\begin{array}}
\def\ea{\end{array}}

\def\beq{\begin{equation}}
\def\eeq{\end{equation}}

\usepackage[normalem]{ulem}
 
\begin{document}

\title{Unravelling the non-Markovian spin-boson model and quantum quasi-Otto cycle}

\author{Shreyas Harshal Pradhan}
\email{shreyas.pradhan@research.iiit.ac.in}
\affiliation{Center for Computational Natural Sciences and Bioinformatics,
International Institute of Information Technology, Hyderabad 500 032, India}

\author{Hadi Mohammed Soufy}
\email{hm.soufy@niser.ac.in}
\affiliation{School of Physical Sciences, National Institute of Science Education and Research, HBNI, Jatni, Bhubaneswar 752 050, India}

\author{Vishal Anand}
\email{vishal.anand@research.iiit.ac.in}
\affiliation{Center for Quantum Science and Technology, International Institute of Information Technology, Hyderabad 500 032, India}
\affiliation{Center for Security, Theory and Algorithmic Research, International Institute of Information Technology, Hyderabad 500 032, India}

\author{Rik Chattopadhyay}
\email{rchattopadhyay.telecom@faculty.iiests.ac.in}
\affiliation{Indian Institute of Engineering Science and Technology, Shibpur, West Bengal 711 103, India}

\author{Subhadip Mitra}
\email{subhadip.mitra@iiit.ac.in}
\affiliation{Center for Computational Natural Sciences and Bioinformatics,
International Institute of Information Technology, Hyderabad 500 032, India}  
\affiliation{Center for Quantum Science and Technology, International Institute of Information Technology, Hyderabad 500 032, India}

\author{Samyadeb Bhattacharya}
\email{samyadeb.b@iiit.ac.in}
\affiliation{Center for Quantum Science and Technology, International Institute of Information Technology, Hyderabad 500 032, India}
\affiliation{Center for Security, Theory and Algorithmic Research, International Institute of Information Technology, Hyderabad 500 032, India}

%%%%%%%%%%%%%%%%%%%%%%%%%%%%%%%%%%%%%%%%%%%%%%%%%%%%%%%%%%%%%%%%%%%%%%%%%%%%%%%%%%%%%%%%%%%%%%%%%%
\begin{abstract}
\noindent We use the spin-boson model to describe the dynamics of a two-level atom interacting with Fabry-P\'erot cavity modes. We solve the Schr\"{o}dinger equation for the system-bath model without the Born-Markov approximation to derive the non-Markovian reduced dynamics of the qubit. We further construct an exact Lindblad-type master equation for it. Similar to the quantum Otto cycle, we construct a non-Markovian quasi-cyclic process based on the atom-cavity interactions, which we call the quasi-Otto cycle. For judicious choices of input state and parameters, the quasi-cycle can be more efficient as a quantum engine than the Otto cycle. We also showed that if the quasi-cycle is repeated multiple times, the efficiency of the quasi-Otto engine asymptotically approaches that of the Otto engine.

%We consider the spin-boson model for a two-level atom interacting with Fabry-P\'erot cavity modes. We solve the Schr\"{o}dinger equation for the system-bath model without considering the Born-Markov approximation to derive a non-Markovian reduced dynamics of the qubit. We further construct an exact Lindblad-type master equation for the same. Our model is applicable not only to cavity QED models, but also in various other scenarios of light-matter interactions or even atom-phonon interactions. We further construct an atom-cavity interaction-based Quantum Otto cycle, which is non-Markovian in nature, to elucidate on its performance compared to the usual Markovian models. Since our model is based on a bird-eye view over experimental setups highly favourable for quantum computational device engineering, the present work has very promising future applications.     
\end{abstract}
%%%%%%%%%%%%%%%%%%%%%%%%%%%%%%%%%%%%%%%%%%%%%%%%%%%%%%%%%%%%%%%%%%%%%%%%%%%%%%%%%%%%%%%%%%%%%%%%%%
\maketitle

%%%%%%%%%%%%%%%%%%%%%%%%%%%%%%%%%%%%%%%%%%%%%%%%%%%%%%%%%%%%%%%%%%%%%%%%%%%%%%%%%%%%%%%%%%%%%%%%%%
\section{Introduction}\label{sec1}
\noindent In the quantum world, physical systems are rarely isolated from the influence of external noise. Systems such as ion traps \citep{ion1,*ion2,*ion3,*ion4,*ion5,*ion6,*ion7}, quantum dots \citep{qd1,*qd2,*qd3,*qd4,*qd5,*qd5,*qd6,*qd7,*qd8}, and Josephson junctions \citep{jos1,*jos2,*jos3} -- known for their promise in various quantum information processing and computational tasks -- are all exposed to environmental interactions, giving rise to dissipation and decoherence. Therefore, one must account for the interaction with environmental degrees of freedom when applying the laws of thermodynamics to a quantum device or system~\citep{otto1,*otto2,*otto3,*otto4,*otto5,*otto6,*otto7,carnot1,*carnot2,*carnot3,*carnot4,*carnot5,*carnot6,*carnot7,*carnot8,*carnot9,*carnot10,*carnot11,ref1,*ref2,*ref3,*ref4}. In general, quantum environments can be of two types -- bosonic and fermionic. In this paper, we consider a bosonic environment and a two-level atomic system coupled to a reflective optical cavity known as the Fabry-P\'erot cavity (FPC)~\citep{fp1}. FPCs~\citep{fp1} have numerous applications in spectroscopy, telecommunication, and sensing~\citep{fp2,*fp3}. With the rapid expansion of integrated optical devices, integrated FPCs are seeing a renewed interest in quantum opto-mechanics. 

In this paper, we construct a realistic theoretical model of the reduced dynamics of the quantum system interacting with the mirror vibration mode(s) of a cavity. The atomic system could be a quantum dot, nano-crystals or ultra-cold atoms like Rubidium. These systems are experimentally realisable and commonly studied in cavity quantum electrodynamics for exploiting controlled light-matter interactions between atomic systems and optical cavities~\citep{cavity1,*cavity2,*cavity3,*cavity4,*cavity5}. (Such controlled interactions play an important role in producing long-range interactions~\citep{cavity6}, constructing photonic gates~\citep{cavity7,*cavity8}, and quantum entanglement-based metrology~\citep{metro1,*metro2,*metro3}.) In the process, we generalise the theoretical paradigm beyond the weak coupling and stationary bath approximation~\citep{book1} to include finite bath models. We also show a thermodynamic device constructed in our setup to demonstrate the possibility of optical cavity-based quantum device engineering with it. 

We start with a simple case, where a qubit, trapped inside a single cavity, interacts with the vibrating mirrors, mimicking a Jaynes-Cummings interaction, and we generalise to the case where the qubit interacts with an arbitrary (but finite) number of vibrational modes close to each other. We derive an exact Lindblad-type master equation describing the qubit dynamics -- non-Markovian by nature -- and analyse its (non-) equilibrium properties. We envisage an experimental scenario where a nano-crystal or a quantum dot sequentially interacts with two multi-mode cavities in thermal equilibrium at different temperatures. With this setup in mind, we design a cavity-based quantum process similar to the Otto Cycle. However, our process differs from that of an ideal Otto engine, where the working medium thermalises with the environment in one crucial aspect. Due to the non-Markovian interaction, which does not allow the system to equilibrate with the mirror modes, our process does not complete a full thermodynamic cycle but a quasi-cycle. Hence, we call it a \emph{quasi}-Otto Cycle. We obtain its efficiency and show that it can be more efficient for some parameter choices than a quantum Otto Cycle.

The plan for the rest of the paper is as follows.
In Section~\ref{dynamics}, we introduce a probable experimental model of the FPC and the theoretical construction. Then we construct the quasi-Otto engine and discuss its properties in Section \ref{sec4}. Finally, we conclude in Section \ref{sec5}.

\section{Dynamics of the two-level system interacting with vibrational modes}\label{dynamics}
\begin{figure}
    \centering
    \includegraphics[width=\columnwidth]{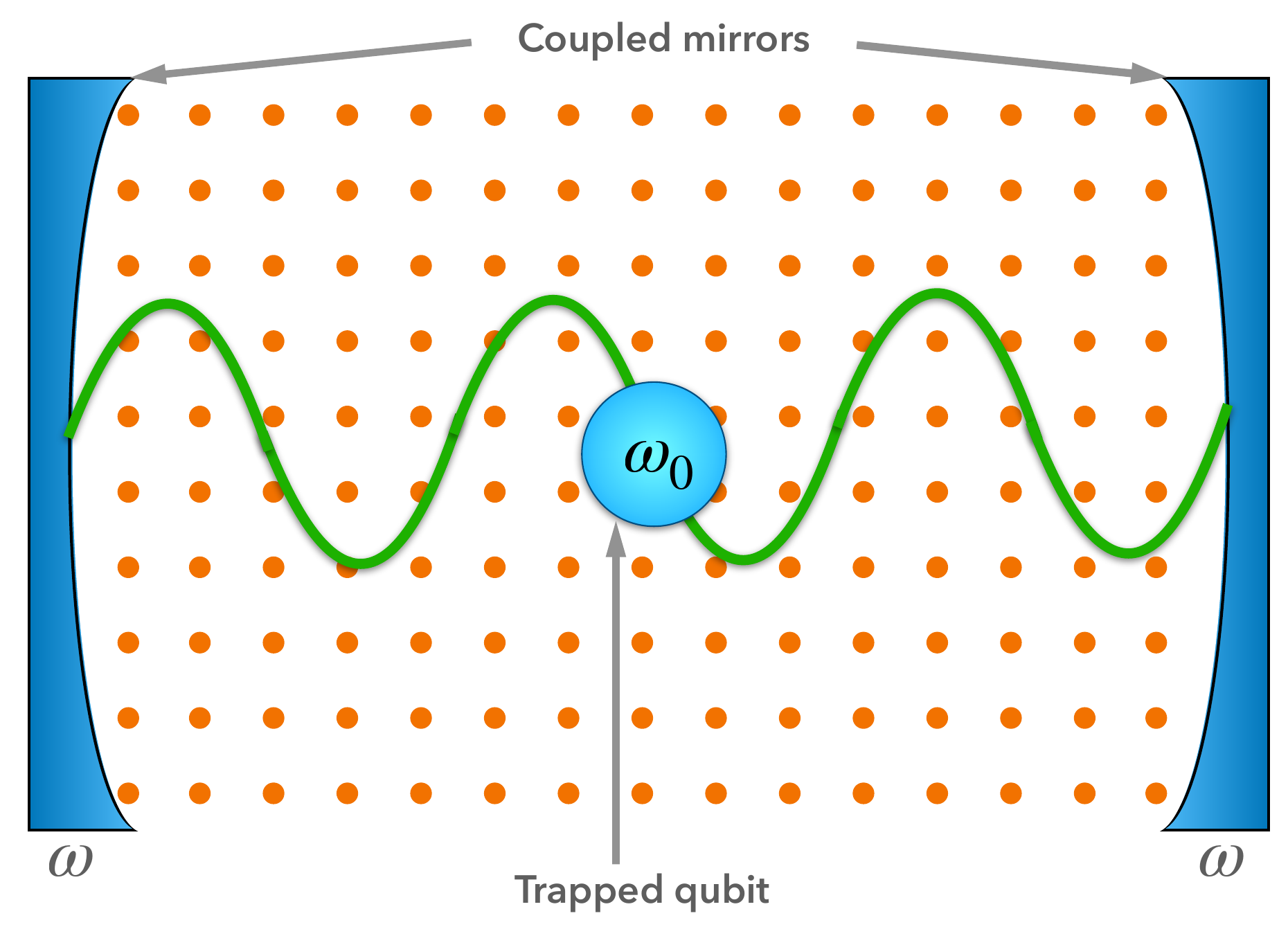}
    \caption{Schematic of the qubit-cavity interaction model.}
    \label{fig:fpc1}
\end{figure}

\noindent
In this section, we introduce the reduced dynamics of a qubit interacting with the vibrational modes of the mirrors. %For motivation, we can consider the following experimental realisation for our theoretical model. %We first describe the experimental model based on which the theory is developed in the following subsection. 
% \subsection{Experimental setup}\label{sec2}
%\noindent 
For an experimental motivation, let us imagine a FPC with two micro-mirrors (Fig.~\ref{fig:fpc1}), which can vibrate at a particular frequency controlled by some external signal. Initially, we assume the two mirrors are tuned to produce a coupled oscillation between them. Then, we introduce a two-level system that emits a photon upon external excitation inside the cavity. By releasing a photon, the qubit excites a trapped optical mode in the cavity that resonates with the vibration mode of the mirrors and, thus, initiates a coupling between the photon and phonons. Now, we can attach the two mirrors to an external thermal bath to observe the dynamics of the qubit in the presence of a thermal bath. Thus, an interaction between the qubit in the cavity and the external bath is established via photon-phonon coupling \citep{CavityExp1,*CavityExp2,*CavityExp3,*CavityExp4,*CavityExp5}. %\textcolor{blue}{Before going into details, we aim to elaborate on the reasoning for the chosen theoretical modelling in the context of the experimental backdrop. Let us consider that the vibrational modes of the mirrors contain many different phononic modes. It is reasonable to consider that the system will interact only one of them, with which it is resonating; i.e. their characteristic frequencies ($\omega_0$ and $\omega$ for the system and vibrational mode respectively) are ideally equal or at most very close to each other with the detuning $|\omega_0-\omega|\ll1$. For this purpose, the dynamics of the system interacting with a single vibrational mode is sufficient for the theoretical description, which we present in the following sub-section. This description is enough to elaborate a one-dimensional cavity, where the electromagnetic wave is reflected between two parallel mirrors placed perpendicularly with the path of wave propagation, as described in Fig. \ref{fig:fpc1}. 
Interestingly, Ref.~\citep{multi-mode} has shown that it is possible that the electromagnetic wave can interact with more than one resonator (vibrating mirror pairs) placed along different directions in the cavity. %\textcolor{red}{For such cases, it is prudent to generalise the dynamics of the system interacting with two or more vibrational modes, which we do in \ref{NmodeModel}}.  % }

%\subsection{Theoretical model}\label{sec3}
%\noindent 
\subsection{Qubit interacting with a single mode}\label{OnemodeModel}
\noindent 
The above scenario can be mapped to a simple Jaynes-Cummings Hamiltonian describing a two-level atom with a transition frequency $\omega_0$, trapped in a cavity with a single mode frequency $\omega$. By adjusting the cavity length, % ($L$),  achieve certain conditions between the system-mode frequency separation, ($|\omega-\omega_0|$), and the axial-mode frequency separation ($c/2L$) can be achieved, where 
we can ensure that the following Hamiltonian describes the qubit-mode interaction:
\begin{equation}\label{1}
    H_T^{(1)} = \hbar\omega_0\sigma_z + \hbar\omega a^\dagger a + \hbar\Delta \left(\sigma_-a^\dagger+\sigma_+ a\right),
\end{equation}
where $H_S=\hbar\omega_0\sigma_z$ and $H_B=\hbar\omega a^\dagger a$ are the self Hamiltonians for the qubit and the bosonic mode, respectively, and $H_{SB}=\hbar\Delta \left(\sigma_-a^\dagger+\sigma_+ a\right)$ is the interaction Hamiltonian. The superscript $(1)$ on the total Hamiltonian, $H_T$, indicates there is only one photon mode. Here, $\sigma_{i}$'s ($i=x,y,z$) are the Pauli matrices, $\sigma_\pm=\sigma_x\pm i\sigma_y$, and $(a^\dagger,~a)$ are the creation and annihilation operators for the mode. The above interaction Hamiltonian follows from the Rabi-oscillation Hamiltonian in the rotating wave approximation~\citep{book1}, where one ignores the counter-rotating terms. This is only possible when $\Delta \ll (\omega,~\omega_0)$. Therefore, with the choice of the Jaynes-Cummings model, we exclude the strong coupling regime from our analysis. 

Without loss of generality, we can restrict ourselves to the following transition,
\begin{align}
\ket{0}\ket{n}\rightarrow&\ A_t^{(1)}\ket{0}\ket{n}+B_t^{(1)}\ket{1}\ket{n-1}=\ket{\psi^{(1)}},\nonumber\\
\ket{1}\ket{n}\rightarrow&\  C_t^{(1)}\ket{1}\ket{n}+D_t^{(1)}\ket{0}\ket{n+1}=\ket{\chi^{(1)}}.\label{2}
\end{align}
We follow the convention: $\ket{0}=(1~0)^T$ and $\ket{1}=(0~1)^T$, representing
the ground and excited states of the qubit, respectively. Here, $\ket{n}$ represents the $n$-th energy eigenstate of the mode. The subscript $t$ on a coefficient indicates its time dependence. The nature of the above state transitions follows from our choice of the Jaynes-Cummings interaction. Under the resonance condition $\omega_0\approx\omega$, the Hamiltonian in Eq.~\eqref{1} is energy conserving, and hence the transitions $\ket{0}\ket{n}\rightarrow\ket{1}\ket{n-1}$ and $\ket{1}\ket{n}\rightarrow\ket{0}\ket{n+1}$ are the only possibilities. %This constraint on the state transition is perfectly legitimate since the interaction Hamiltonian is tailor-made for such energy exchanges. 
Considering the Schr\"{o}dinger equation 
\begin{equation*}
i\hbar\frac{d}{dt}\ket{\psi\textcolor{blue}{^{(1)}}} = H_T\ket{\psi\textcolor{blue}{^{(1)}}},    
\end{equation*}
we obtain the following set of differential equations,
\begin{align}
\dot{A}_t^{(1)} =&\ -i(\omega_0+n\omega)A_t^{(1)}-i\Delta\sqrt{n}B_t^{(1)},\nonumber\\
\dot{B}_t^{(1)} =&\ i(\omega_0+(n-1)\omega)B_t^{(1)}-i\Delta\sqrt{n}A_t^{(1)}. \label{3}
\end{align}
Solving these coupled differential equations, we get
\begin{align}
    A_t^{(1)} =&\ e^{-i\alpha^{(1)} t/2}\left(\cos{\left(\eta^{(1)} t/2\right)}-i\frac{\beta^{(1)}}{\eta^{(1)}}\sin{\left(\eta^{(1)} t/2\right)}\right), \nonumber\\
    B_t^{(1)} =&\ -2i\Delta\sqrt{n}~e^{-i\alpha^{(1)} t/2}\left(\frac{\sin{\left(\eta^{(1)} t/2\right)}}{\eta^{(1)}}\right),
\label{4}
\end{align}
where
\begin{align*}
\alpha^{(1)} =&\ (2n-1)\omega,~~\beta^{(1)} = 2\omega_0+\omega, {\rm ~and}\\
 \eta^{(1)} =&\ \sqrt{(\beta^{(1)})^2+4n\Delta^2}.
\end{align*}
It is straightforward to check that 
\[
|A_t^{(1)}|^2 = 1 - |B_t^{(1)}|^2.
\]
Similarly, solving the Schr\"{o}dinger equation for the state $\ket{\chi^{(1)}}$, we get 
\begin{align}
    C_t^{(1)} =&\ e^{-i\alpha'^{(1)} t/2}\left(\cos{\left(\eta'^{(1)} t/2\right)}-i\frac{\beta'^{(1)}}{\eta'^{(1)}}\sin{\left(\eta'^{(1)} t/2\right)}\right)\nonumber\\
    D_t =&\ -2i\Delta\sqrt{n+1}~e^{-i\alpha'^{(1)} t/2}\left(\frac{\sin{\left(\eta'^{(1)} t/2\right)}}{\eta'^{(1)}}\right),
\label{5}
\end{align}
where
\begin{align*}
\alpha'^{(1)} =&\ (2n+1)\omega,~~\beta'^{(1)} = -( 2\omega_0+\omega), {\rm ~and}\\
 \eta'^{(1)} =&\ \sqrt{(\beta'^{(1)})^2+4(n+1)\Delta^2}.
\end{align*}
The coefficients obey the following relation:
\[
|C_t^{(1)}|^2=1-|D_t^{(1)}|^2.
\]
To obtain the reduced dynamics of the qubit, we consider the initial system-bath state $\rho_s\otimes\tau_\gamma$, where $\rho_s=\left(\begin{matrix}
\rho_{00} & \rho_{01}\\
\rho_{10} & \rho_{11}
\end{matrix}\right)$ is an arbitrary system state and $\tau_\gamma = \exp{(-\gamma H_B)}/\mathcal{Z}$ is the thermal bath state ($\gamma=1/k_BT$ is the inverse temperature and $\mathcal{Z}$ is the partition function). We get the components of the reduced density matrix by tracing over the bath modes:
\begin{align*}
{\rm Tr}_B\left[\ket{\psi^{(1)}}\bra{\psi^{(1)}}\right] =&\  |A_t^{(1)}|^2\ket{0}\bra{0}+|B_t^{(1)}|^2\ket{1}\bra{1},\\
{\rm Tr}_B\left[\ket{\chi^{(1)}}\bra{\chi^{(1)}}\right] =&\  |D_t^{(1)}|^2\ket{0}\bra{0}+|C_t^{(1)}|^2\ket{1}\bra{1},\\
{\rm Tr}_B\left[\ket{\psi^{(1)}}\bra{\chi^{(1)}}\right] =&\ (A_t^{(1)})(C^{(1)}_t)^*\ket{0}\bra{1}.    
\end{align*}
With the above relations, we can express the dynamical map of the reduced density matrix as
\begin{align}
\rho_{00}(t) =&\ (1-\mathcal{A}_t^{(1)})~\rho_{00}+\mathcal{B}_t^{(1)}~\rho_{11},\nonumber\\
\rho_{11}(t) =&\ (1-\mathcal{B}_t^{(1)})~\rho_{11}+\mathcal{A}_t^{(1)}~\rho_{00},\nonumber\\
\rho_{01}(t) =&\ \mathcal{C}_t^{(1)}~\rho_{01},~~\rho_{10}(t)=\rho_{01}^*(t),
\label{6}
\end{align}
with 
\begin{align}
\mathcal{A}_t^{(1)} =&\ \frac{1}{\mathcal{Z}}\sum_{n=0}^\infty 4n\Delta^2 \left(\frac{\sin{\left(\eta^{(1)} t/2\right)}}{\eta^{(1)}}\right)^2~\exp{(-\gamma n\hbar\omega}),\nonumber\\
\mathcal{B}_t^{(1)} =&\ \frac{1}{\mathcal{Z}}\sum_{n=0}^\infty 4(n+1)\Delta^2\left(\frac{\sin{\left(\eta'^{(1)} t/2\right)}}{\eta'^{(1)}}\right)^2~\exp{(-\gamma n\hbar\omega}),\nonumber\\
\mathcal{C}_t^{(1)} =&\ \frac{1}{\mathcal{Z}}\sum_{n=0}^\infty e^{-i\omega t}~\left(\cos{\left(\frac{\eta^{(1)} t}2\right)}-i\frac{\beta^{(1)}}{\eta^{(1)}}\sin{\left(\frac{\eta^{(1)} t}2\right)}\right)\nonumber\\
&\hspace{-0.5cm} \times \left(\cos{\left(\frac{\eta'^{(1)} t}2\right)}+i\frac{\beta'^{(1)}}{\eta'^{(1)}}\sin{\left(\frac{\eta'^{(1)} t}2\right)}\right)~\exp{(-\gamma n\hbar\omega}).
\label{7}
\end{align}
%Here, `$t$' in the parenthesis represents the time dependence of the terms.

\subsection{Qubit interacting with multiple modes}\label{NmodeModel}
\noindent 
We can easily generalise the above analysis of a two-level atom interacting with a single mode to a qubit system interacting with multiple close modes that do not directly interact with each other. As mentioned earlier, this kind of situation may arise when the system interacts with two or more resonators in the cavity. The theoretical inspiration for this model stems from the study of Markovian spin-Boson interaction, where the environment is assumed to be a static collection of an infinite number of Harmonic oscillators so that the Born-Markov approximation can be used to get the reduced system dynamics described by a Markovian Lindblad master equation~\citep{book1}. We generalise the situation by considering the system qubit interacting with a finite number of modes that need not be static collectively, and do not use the Born-Markov approximation. Thus, the dynamics generalises to that of a system interacting with a finite number of bath modes under an essentially non-Markovian paradigm. 
%A more apt and system specific description is that the two level system is interacting with multiple coupled cavity micro mirrors, described by finite number of phononic modes. 

%Our theoretical inspiration for this model stems from the study of Markovian spin-Boson interaction, where the environment is assumed to be a static collection of infinite number of Harmonic oscillators considered to be a static environment, upon which Born-Markov approximation is imposed to get the reduced system dynamics described by a Markovian Lindblad master equation \citep{book1}. In our analysis, we generalise the situation by relaxing most of the primary assumptions. We consider that the system qubit is interacting with finite number of modes, the collection of which is not static under the interaction. Moreover, we also do not consider the Born-Markov approximation \citep{book1}. This generalises our dynamics to the case where a system is interacting with a finite number of bath modes under essentially non-Markovian paradigm. A more apt and system specific description is that the two level system is interacting with multiple coupled cavity micro mirrors, described by finite number of phononic modes. 
%It is to be noted that the basic theoretical description can be adopted to describe many other experimentally feasible scenario, where a system is interacting with a finite number of bosonic modes. 

For the above setup, the total Hamiltonian of the qubit interacting with an arbitrary number of cavity modes can be described as an extension of the Jaynes-Cummings interaction as follows:
\begin{align}
H_T^{(N)} =&\ \hbar\omega_0\sigma_z + \hbar \sum_{i=1}^N \omega_ia_i^\dagger a_i + \hbar\Delta \sum_{i=1}^N \big(\sigma_-a_i^\dagger+\sigma_+ a_i\big)\nonumber\\
\approx&\ \hbar\omega_0\sigma_z + \hbar \omega\sum_{i=1}^N a_i^\dagger a_i + \hbar\Delta \sum_{i=1}^N \big(\sigma_-a_i^\dagger+\sigma_+ a_i\big).\label{15}
\end{align}
Here, we have ignored the minor differences between the $\omega_i$'s in the second step as the modes are close and set $\omega_i\approx \omega$ for all $i=[1, N]$ as a first approximation for simplicity and to get a compact analytic expression. The only physical constraint we impose is that the interaction strength ($\Delta$) is suppressed compared to the free Hamiltonian of the modes, i.e., $\Delta \ll \omega$. These considerations allow us to express the system-bath transitions similar to the previous case:
\begin{align}
&\hspace{-0.4cm}\ket{0}\ket{n_1}\cdots\ket{n_N}\rightarrow\nonumber\\
&A_t^{(N)}\ket{0}\ket{n_1}\cdots\ket{n_N}+\sum_{i=1}^N B_{i,t}^{(N)}\ket{1}\ket{n_1-\delta_{i1}}\cdots\ket{n_N-\delta_{iN}},\nonumber\\
&\hspace{-0.4cm}\ket{1}\ket{n_1}\cdots\ket{n_N}\rightarrow\nonumber\\
&C_t^{(N)}\ket{1}\ket{n_1}\cdots\ket{n_N}+\sum_{i=1}^N D_{i,t}^{(N)}\ket{0}\ket{n_1+\delta_{i1}}\cdots\ket{n_N+\delta_{iN}},\label{16}
\end{align}
where $\delta_{ij}$ is the Kronecker delta function. These transitions correspond to the solutions of the total system-bath Schr\"{o}dinger equation governed by the Hamiltonian in Eq.~\eqref{15}, as long as transitions between different bath modes are not allowed. Now, since $\Delta \ll \omega$, the modes oscillate much faster than the system-mode interactions, which are responsible for the system-bath and bath-bath (via the system) transition. As a result, bath-to-bath transitions and, consequently, bath-bath correlations become highly improbable.  
%Hence, we can neglect them in comparison to other allowed transitions. But, since $\omega_0$ and $\Delta$ are of comparable order of magnitude, system-bath interaction cannot be ignored from the perspective of the qubit evolution. 
Under this physical constraint, the transitions represented in Eq.~\eqref{16} are the only feasible possibilities and hence, the qubit evolution can be described as 
\begin{align}
\rho_{00}(t) =&\ (1-\mathcal{A}_t^{(N)})~\rho_{00}+\mathcal{B}_t^{(N)}~\rho_{11},\nonumber\\
\rho_{11}(t) =&\ (1-\mathcal{B}_t^{(N)})~\rho_{11}+\mathcal{A}_t^{(N)}~\rho_{00},\nonumber\\
\rho_{01}(t) =&\ \mathcal{C}_t^{(N)}~\rho_{01},~~\rho_{10}(t)=\rho_{01}^*(t),\label{17}
\end{align}
with 
\begin{align}
\mathcal{A}_t^{(N)} =&\ \frac{1}{\Pi_{i=1}^N\mathcal{Z}_i}\mathlarger{\sum_{n_{tot}=0}^\infty}\ ^{(N+n_{tot}-1)}C_{n_{tot}} \times 4n_{tot}\Delta^2 \nonumber\\    
&\hspace{0.2cm}\times\Bigg(\frac{\sin{\left(\eta^{(N)} t/2\right)}}{\eta^{(N)}}\Bigg)^2\exp{(-\gamma\, n_{tot}\hbar\omega}),\nonumber\\
\mathcal{B}_t^{(N)} =&\ \frac{1}{\Pi_{i=1}^N\mathcal{Z}_i}\mathlarger{\sum_{n_{tot}=0}^\infty}\ ^{(N+n_{tot}-1)}C_{n_{tot}} \times 4(n_{tot}+1)\Delta^2\nonumber\\
&\hspace{0.2cm}\times\Bigg(\frac{\sin{\left(\eta'^{(N)} t/2\right)}}{\eta'^{(N)}}\Bigg)^2\exp{(-\gamma\, n_{tot}\hbar\omega}),\nonumber\\
\mathcal{C}_t^{(N)} =&\ \frac{e^{-i\omega t}}{\Pi_{i=1}^N\mathcal{Z}_i}\mathlarger{\sum_{n_{tot}=0}^\infty} \ ^{(N+n_{tot}-1)}C_{n_{tot}} \exp{(-\gamma\, n_{tot}\hbar\omega)} \nonumber\\
&\hspace{0.2cm}\times\Bigg(\cos{\left(\eta^{(N)} t/2\right)}-i\frac{\beta^{(N)}}{\eta^{(N)}}\sin{\left(\eta^{(N)} t/2\right)}\Bigg)\nonumber\\
&\hspace{0.2cm}\times\Bigg(\cos{\left(\eta'^{(N)} t/2\right)}-i\frac{\beta^{(N)}}{\eta'^{(N)}}\sin{\left(\eta'^{(N)} t/2\right)}\Bigg),\label{18}
\end{align}
where 
\[
\begin{array}{ll}
n_{tot} = n_1+n_2+\cdots+n_N,\\~\\
\alpha^{(N)} = \left(2 n_{tot} -1 \right)\omega, \alpha'^{(N)} = \left(2n_{tot} +1 \right)\omega,\\
\\
\beta^{N} = 2\omega_0 + \omega , ~\eta^{(N)} = \sqrt{(\beta^{N})^2+4n_{tot}\Delta^2},\\
%\\
%\beta'^{(N)} = -(2\omega_0 + \omega) ,\\
\\
\eta'^{(N)} = \sqrt{(\beta^{(N)})^2+4(n_{tot}+N)\Delta^2}.
\end{array}
\]
We derive the above expressions by first considering the $N=2$ and $3$ cases explicitly, and then generalising. See Appendix~\hyperref[appA]{A} for the $N=2$ case. We also construct the exact Lindblad-type master equation for this dynamics in Appendix~\hyperref[appB]{B}.

\subsection{Non-equilibrium nature of the qubit evolution} 
\noindent %Before proceeding to the construction of a thermodynamic cycle we 
Let us now investigate the thermalisation property of the evolution in Eq.~\eqref{17}. When a comparatively smaller system interacts with a thermal bath, it is expected to come to thermal equilibrium with the bath under ideal circumstances. But, due to the non-Markovian nature of the system-bath interaction~\citep{nm1,*nm2,*nm3,*nm4,*nm5,*nm7,*nm8,*nm9,*nm10,*nm11,nm6}, there can be information backflow~\citep{nm6} driving the system out of equilibrium. To investigate whether the system thermalises or not after a sufficiently long time, we consider the long-time averaged state $\overline{\rho} = \lim_{T\rightarrow\infty}\int_0^t\rho (t)dt/T$. In a Markovian evolution, where there is no information backflow, the time-averaged state should be the thermal state~\citep{ergo}. This can be easily demonstrated by considering the following Markovian evolution: $\rho(t)= \rho(0)e^{-at}+\tau_\gamma (1-e^{-at})$, where $\tau_\gamma$ is the thermal state. %For this evolution, it is straightforward to check that $\overline{\rho}=\tau_\gamma$. 
To verify whether the evolution given in Eq.~\eqref{17} thermalises, we consider the optimised trace distance calculated as 
\[
D(\overline{\rho},\tau_\gamma)=\min_{\rho(0)}\frac{1}{2}||\overline{\rho}-\tau_\gamma||_1= \frac{1}{2}|\chi^{(N)}-\overline{\mathcal{A}}^{(N)}-\tanh{\gamma\hbar\omega_0}|.
\] 
Here $\overline{\mathcal{A}}^{(N)}(\chi^{(N)})=\lim_{T\rightarrow\infty}\int_0^T\mathcal{A}_t^{(N)}(\mathcal{B}_t^{(N)})dt/T$. This means that the system thermalises if the following condition holds:
\begin{equation}
    \mathcal R= \frac{1}{2\gamma\hbar\omega_0}\ln{\left(\frac{1+(\chi_t^{(N))}-\overline{\mathcal{A}}^{(N)})}{1-(\chi^{(N)}-\overline{\mathcal{A}}^{(N)})}\right)}=1.\label{eq:ratioR}
\end{equation}

\begin{figure}[!b]
    \centering
    \includegraphics[width=\columnwidth]{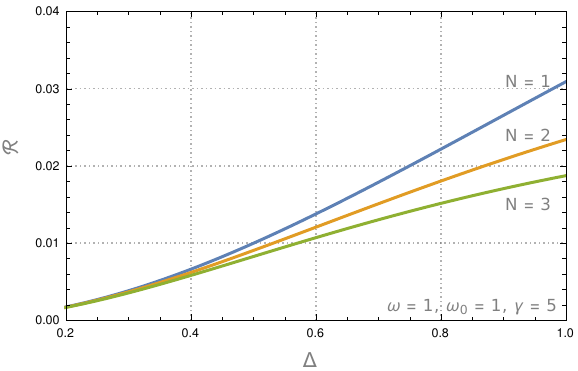}
    \caption{The ratio $\mathcal{R}$, defined in Eq.~\eqref{eq:ratioR}, for different $N$ values as functions of $\Delta$, the system-mode ($\equiv$ system-bath) coupling. Clearly, since $\mathcal{R}\neq1$, the system never thermalises for $\Delta<\omega$. Here, we consider the natural unit system and set $\hbar=k_B=1$.}
    \label{fig:eqilib}
\end{figure}

In Fig. \ref{fig:eqilib}, we show how the ratio $\mathcal{R}$ varies with $\Delta$ for different numbers of modes, $N$ and a typical parameter choice. The system does not thermalise in these cases, and thus exhibits a non-equilibrium phenomenon. (In fact, $\mathcal R$ remains much smaller than $1$ for these parameters, even if we take $\Delta>1$, beyond the region of our interest.) The reason for this behaviour can be interpreted as a consequence of the memory effect induced by the non-Markovian nature of the system-mode (bath) interaction. We will later make use of this interesting characteristic in our construction of the `quasi-Otto cycle'.

 \section{Quantum quasi-Otto cycle}\label{sec4}
\noindent
In this section, we demonstrate an experimentally relevant application of our theoretical construction: a quantum thermodynamic quasi-cyclic process. For this purpose, we first consider a cavity-based toy model of the quantum Otto cycle~\citep{otto1,otto2,otto3,otto4,otto5,otto6,otto7}, where the working medium is a qubit coupled to two heat baths at temperatures $T_c$ and $T_h$ ($T_c<T_h$), called cold and hot baths, respectively. 
%The dynamical model presented in the previous section, allows us to consider finite time thermodynamic ``quasi-cyclic" processes, closely mimicking its ``perfectly cyclic" counterparts. 
In recent years, novel experimental techniques have been developed to access both weak and strong coupling regimes~\citep{expt1,*expt2,*expt3,*expt4,*expt5}. The theory side also has seen similar progress in studying the performance of thermal machines beyond the weak coupling scenario~\citep{expt6,*expt7,*expt8}. In this backdrop, we construct a non-equilibrium model of a quantum quasi-cyclic process. The quasi-cyclic nature is due to the non-equilibrium nature of the interaction: the system does not thermalise and hence cannot complete a perfect cycle. However, before discussing the process, we quickly look at the description of heat, work and energy conservation principle at the quantum level.

The change in the total energy of a quantum system is given as: ${\rm Tr}[H_i\rho_i]-{\rm Tr}[H_f\rho_f]$, where $(H_i,~\rho_i)$ and $(H_f,~\rho_f)$ represent the (initial Hamiltonian, initial state) and the (final Hamiltonian, final state) pairs, respectively. Therefore, for a time-dependent process, the rate of change of average energy can be written as: 
\[
\frac{d}{dt}\langle H(t)\rangle = {\rm Tr}\left[\dot{H}(t)\rho(t)\right] + {\rm Tr}\left[H(t)\dot{\rho}(t)\right].
\]
The first term on the right-hand side is the average power $\mathcal{P}(t)=-{\rm Tr}\left[\dot{H}(t)\rho(t)\right]$
and the second term is the rate of change of average heat intake $\frac{d}{dt}Q(t)=-{\rm Tr}\left[H(t)\dot{\rho}(t)\right]$ with time. Hence, we can express the first law of thermodynamics as,
$
\Delta E=Q + \mathcal{W},
$
where $Q=\int_{t_i}^{t_f}\frac{d}{dt}Q(t)dt$ is the total heat exchange and $\mathcal{W}=\int_{t_i}^{t_f}\mathcal{P}(t)dt$ is the total work done by the system.

\subsection{The quantum Otto cycle}
\begin{figure}[!t]
    \centering
    \includegraphics[width=\columnwidth]{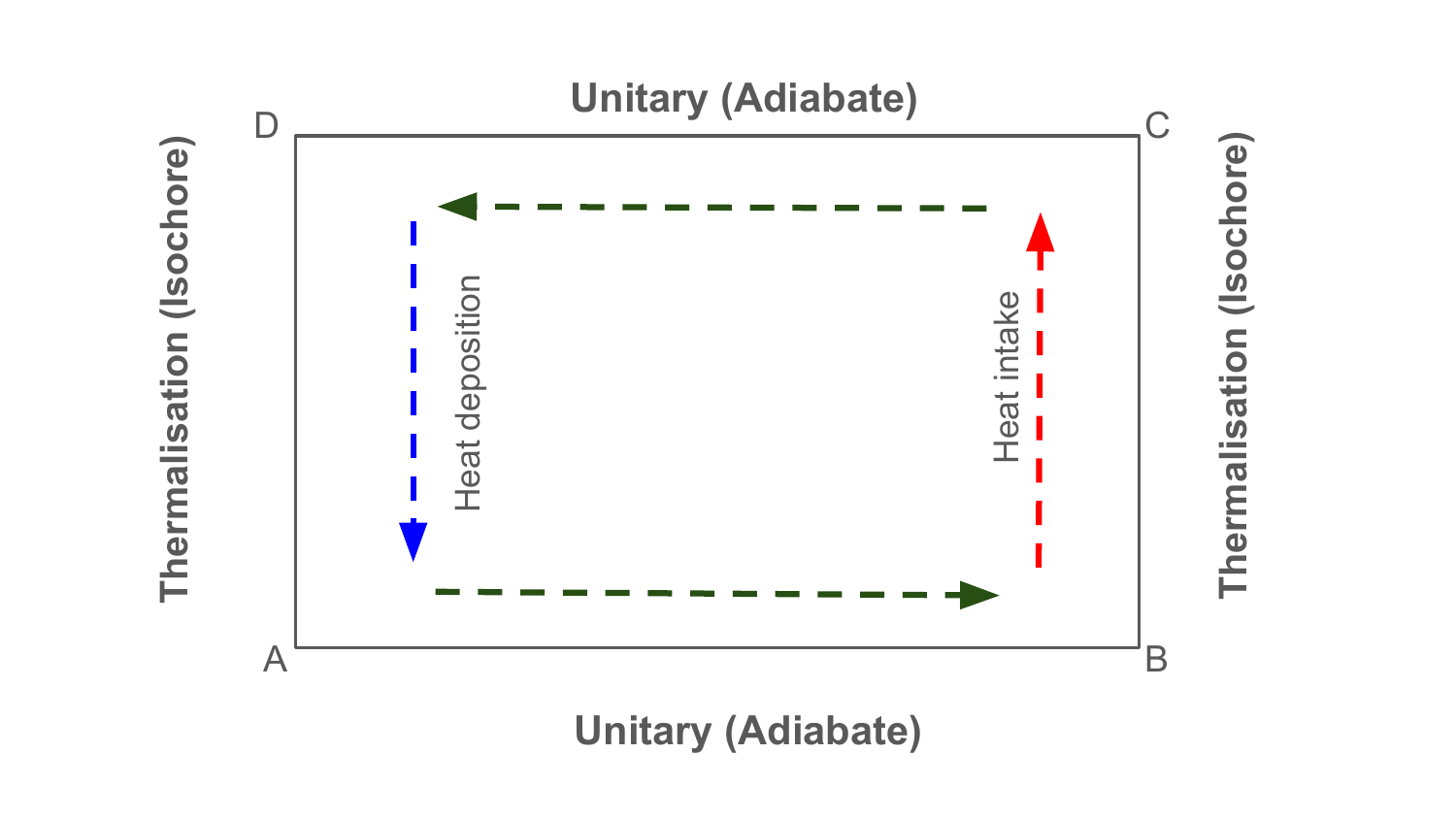}
    \caption{The quantum Otto cycle.}
    \label{fig:ottoPic}
\end{figure}

\noindent
The ideal Otto cycle is made up of alternating adiabatic and isochoric strokes. We show the schematic of the cycle in Fig.~\ref{fig:ottoPic}. In the quantum version, the first stroke is a unitary process mimicking an adiabatic process. Initially, the qubit is at thermal equilibrium with the cold bath, i.e., it is in the thermal state at temperature $T_c$, i.e., $\rho_s = \exp{(-\gamma_cH^c_S)/{\rm Tr}[\exp{(-\gamma_cH^c_S)}]}$, with $\gamma_c=1/k_BT_c$ and the system frequency, $\omega_0=\omega_c$. The unitary stroke drives the system via a driving Hamiltonian $H(t)=\hbar\omega(t)\sigma_Z$, where its frequency changes from $\omega_c$ to $\omega_h$. In the second stroke, the qubit is connected to the hot bath at temperature $T_h$ and kept for a sufficient time for it to thermalise with the bath.
%consisting of a number of cavities considered to be in a thermal state with temperature $T_h$ (inverse temperature $\gamma_h=1/K_BT_h$) to let it thermalize. After a sufficiently long time, we disconnect the qubit from the bath, completing the second stroke. Then as a third stroke, 
In the third stroke, a second unitary drives the system to change its frequency from $\omega_h$ to $\omega_c$. In the final stroke of the process, the system thermalises with the cold bath. If $Q_h$ is the amount of heat extracted from the hot bath and $Q_c$ is the amount ($|Q_c| <|Q_h|$) dumped into the cold bath, we can calculate the efficiency of the Otto engine as~\citep{otto1,otto2,otto3,otto4},
\begin{equation}
    \mathcal{E}_O=1-\frac{|Q_c|}{|Q_h|}=1-\frac{\omega_c}{\omega_h}.
\end{equation}
The second law of thermodynamics prevents any heat engine from being more efficient than the Carnot engine. Thus, we have
\begin{align}
  \mathcal{E}_O=1-\frac{\omega_c}{\omega_h} \leq \mathcal{E}_C = 1-\frac{\gamma_h}{\gamma_c}.\label{eq:ottocarnot}  
\end{align}

\subsection{Construction of the quasi-Otto cycle}
\noindent
As we saw in the previous section, the qubit need not be thermalised in our model even after interacting with a bath (via a cavity mode) for a long time. Hence, in such cases, it is impossible to realise the ideal Otto cycle in our setup, even if we arrange the qubit to interact with a pair of hot and cold baths cyclically. Nevertheless, we can have a similar `periodic' process. However, first, let us examine how this process can be realised experimentally..

\begin{figure}[!t]
    \centering
    \includegraphics[width=\columnwidth]{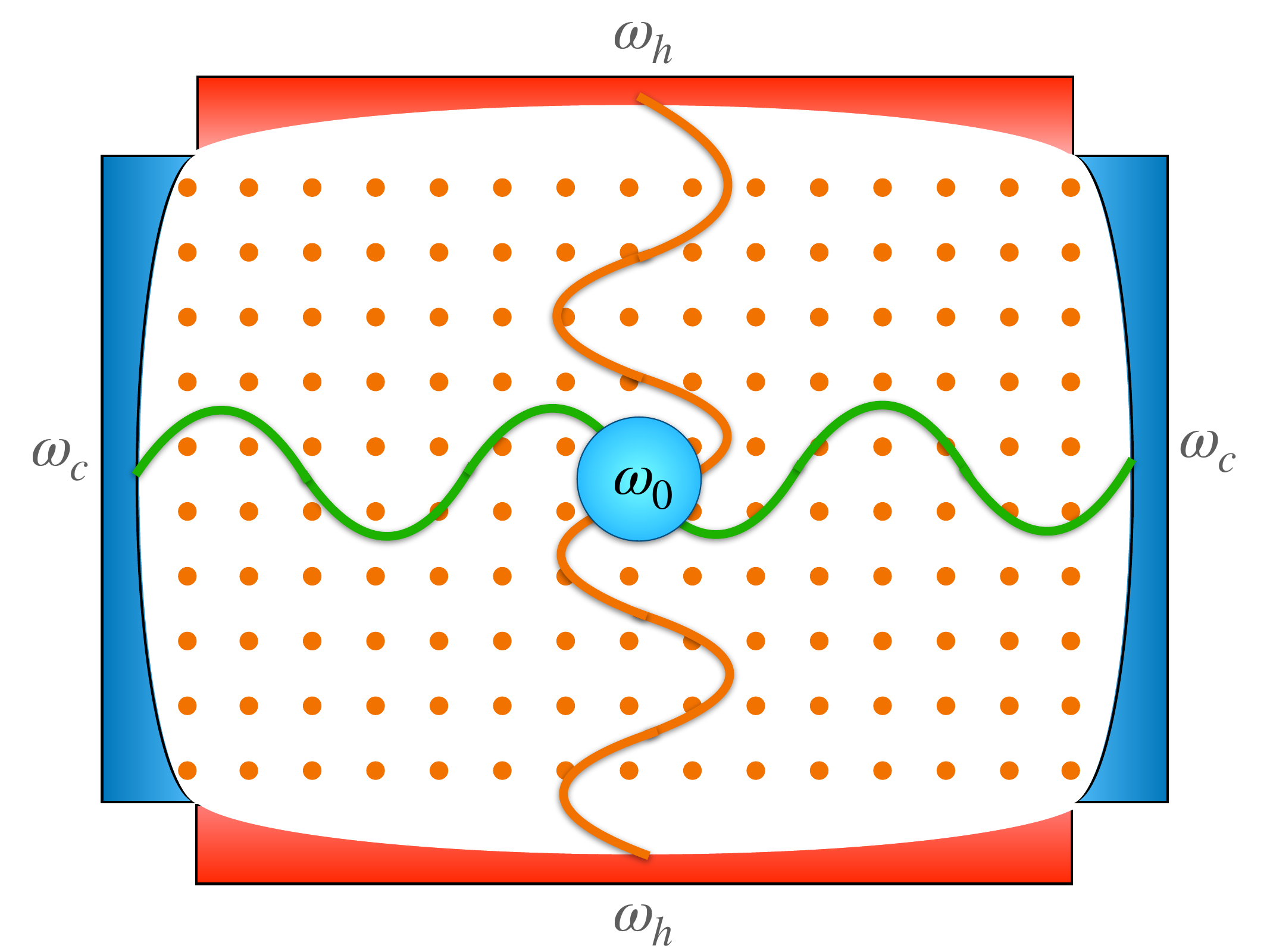}
    \caption{Schematic diagram of the experimental setup for the quasi-Otto engine.}
    \label{fig:otto}
\end{figure}

Let us imagine a modified FPC that combines two FPCs with two orthogonally arranged micro mirror pairs as illustrated in Fig.~\ref{fig:otto}. Each pair is linked to a heat bath, one being hotter than the other. Consequently, the resonant modes of these mirror pairs engage with hot and cold baths. Let the frequencies of the resonant modes supported by these two mirror pairs be $\omega_h$ and $\omega_c$, respectively. When an external optical pump is applied, the pair connected to the hot bath (marked with red in the figure) stimulates a photonic mode of frequency $\omega_h$ inside the cavity (indicated by the vertical orange wavy line), allowing the qubit to absorb the photon and transit to the excited state. Conversely, as the qubit returns to the ground state, it emits a photon at $\omega_c$, activating the photonic cavity mode (shown as the horizontal green wavy line) that resonates with the cold bath mirrors (shown as blue). Depending on its energy states, this configuration allows the qubit to interact with the exterior thermal bath through photon-phonon coupling. With this setup in mind, we can now construct the process.

\begin{enumerate}
    \item \textbf{First stroke:} As in the case of the ideal Otto cycle, let us start our cycle when $\omega_0=\omega_c$. In principle, we could consider the working medium to be in contact with the cold bath and be in thermal equilibrium before we isolate it and start the first unitary evolution corresponding to the Hamiltonian $H(t)=\hbar\omega(t)\sigma_z$ where $\omega(t_i)=\omega_c$ and $\omega(t_f)=\omega_h$. However, for convenience, we consider a general initial state $\rho_1=x_1\ket{0}\bra{0}+(1-x_1)\ket{1}\bra{1}$. This unitary process can be understood as the optical pump stimulating the system to attain the frequency of vibration of the hot mirror pair to resonate and thus interact with the hot bath. The work done by this unitary process is given by 
\[
W_1 = \hbar(\omega_c-\omega_h)(2x_1-1).
\]

    \item 
    \textbf{Second stroke:} The second step is the first isochore, where the qubit interacts with the hot bath. The non-Markovian nature of the system-bath interaction implies that the system may not come to thermal equilibrium with the bath in a reasonable period. Therefore, we allow it to be in thermal contact for a sufficiently long time ($t^h_{i}\to t^h_{f}$) and then isolate it from the bath. This is when the system is in resonance with the hot mirror and thus interacts with it to exchange energy. Let us consider that the system reaches a state $\rho_2=y_1\ket{0}\bra{0}+(1-y_1)\ket{1}\bra{1}$ after being in resonance with the hot mirror for a sufficiently long time. The energy exchange can, therefore, be calculated as 
\[
\Delta E_h = 2\hbar\omega_h(y_1-x_1),
\]
of which
\[
Q_h=-\int_{t^h_{i}}^{t^h_{f}}{\rm Tr}[U_h(t)\sigma_z\dot{\rho}(t)dt]
\] 
is the heat exchange and 
\[
\mathcal{W}_h=-\int_{t^h_{i}}^{t^h_{f}}{\rm Tr}[\dot{U}_h(t)\sigma_z\rho(t)dt]
\]
is the internal work done by the time-dependent self-Hamiltonian $U_h(t)\sigma_z$. This is interesting because, in this case, the thermal interaction with the hot bath not only gives rise to a heat exchange but also does some work. %This excess work done appears due to the time-dependent self-Hamiltonian $U_h(t)\sigma_z$ generated by the system-bath interaction.

\item 
\textbf{Third stroke:} Similarly, for the reverse adiabatic process bringing the characteristic frequency of the system from $\omega_h$ to $\omega_c$, the work done is 
\[
W_2 = \hbar(\omega_h-\omega_c)(2y_1-1).
\]
This happens when the optical pumping is applied again to bring the system in resonance with the cold mirror vibration.

\item 
\textbf{Forth stroke:} In the fourth and final stroke of the process, the system in state $\rho_2$ interacts with the cold bath for a sufficiently long period ($t^c_i\to t^c_f$). For this process, the total change in the energy of the system is 
\[
\Delta E_c = 2\hbar\omega_c(y_1-x_2),\\
\]
where $\rho_3=x_2\ket{0}\bra{0}+(1-x_2)\ket{1}\bra{1}$ is the arbitrary final state after we let $\rho_2$ interact with the cold bath. Here, this energy exchange is divided into
\[
\mathcal{W}_c=-{\rm Tr}\left[\int_{t^c_{i}}^{t^c_{f}}\dot{U}_c(t)\sigma_z\rho(t)dt\right],
\]  
the intrinsic work done, and 
\[
Q_c=-Tr\left[\int_{t^c_{i}}^{t^c_{f}}U_c(t)\sigma_z\dot{\rho}(t)dt\right],
\]
the heat exchange.
\end{enumerate}

\subsection{Efficiency of the quasi-Otto engine: Single run} 
\noindent
Before considering the quasi-Otto cycle as an engine, we consider two points. First, the quantum Otto engine's efficiency depends solely on the heat exchanges with the hot and cold baths, as there is no time-dependent driving Hamiltonian for these thermal interactions. However, in our case, we have the Hamiltonians, $U_h(t)\sigma_z$ and $U_c(t)\sigma_z$, giving rise to the intrinsic works, $\mathcal{W}_h$ and $\mathcal{W}_c$, respectively, which can either help or hinder the engine. Hence, we use the total energy exchange instead of the heat transfer to estimate the efficiency by accounting for these. We calculate the efficiency of the quasi-Otto cycle as,
\begin{equation}
    \mathcal{E}_{qO} = \frac{|\Delta E_h|-|\Delta E_c|}{|\Delta E_h|} = 1-\frac{\omega_c}{\omega_h}\left|\frac{y_1-x_2}{y_1-x_1}\right|.\label{eff1}
\end{equation}
%Clearly, that ideal Otto efficiency gets modified by a multiplicating factor, which depends on the components of the target states. Note that 
If $x_2=x_1$, i.e., if the cycle completes perfectly, we get back the Otto efficiency $\mathcal{E}_O$. It is also clear from Eq.~\eqref{eq:ottocarnot} that, depending on $x_1$, $x_2$, and $y_1$, $\mathcal{E}_{qO}$ can be both larger or smaller than $\mathcal{E}_{O}$. However, since it is always bounded by $\mathcal{E}_{C}$, we have
\begin{equation}\label{carnot-bound}
\frac{\gamma_h\omega_h}{\gamma_c\omega_c}\leq \left|\frac{y_1-x_2}{y_1-x_1}\right|.
\end{equation}
A violation of Eq.~\eqref{carnot-bound} implies that the quasi-cycle cannot work as a physical heat engine. So, as a physical quantum engine, the quasi-Otto cycle performs better than the ideal Otto engine when
\begin{equation}\label{moreEff}
    \frac{\gamma_h\omega_h}{\gamma_c\omega_c}\leq\left|\frac{y_1-x_2}{y_1-x_1}\right|\leq 1.
\end{equation}
If, however, the Otto cycle converges to the Carnot cycle, i.e., $\omega_c/\omega_h=\gamma_h/\gamma_c$, then $\mathcal{E}_{qO}\leq \mathcal{E}_{O}$ and we get $x_1\geq x_2$.

Second, the qubit interacts for long stretches of time in the second and fourth strokes of the quasi-Otto cycle. Hence, for practical purposes, we can consider the long-time averaged states as the target states of these strokes: 
\begin{align*}
\langle\rho\rangle_\infty = \lim_{T\rightarrow\infty}\frac{1}{T}\int_0^T \rho_t dt.     
\end{align*}
Thus, we get, 
\begin{align}
    y_1 =&\ (1-\overline{\mathcal{A}}_h^{(N)})x_1+\overline{\mathcal B}_h^{(N)}(1-x_1),\nonumber\\
    x_2 =&\ (1-\overline{\mathcal{A}}_c^{(N)})y_1+\overline{\mathcal B}_c^{(N)}(1-y_1)\nonumber\\
        =&\ [(1-\overline{\mathcal{A}}_c^{(N)})(1-\overline{\mathcal{A}}_h^{(N)})+\overline{\mathcal{A}}_h^{(N)}\overline{\mathcal B}_c^{(N)}]x_1\nonumber\\
    &\ +[\overline{\mathcal B}_h^{(N)}(1-\overline{\mathcal{A}}_c^{(N)})+\overline{\mathcal B}_c^{(N)}(1-\overline{\mathcal B}_h^{(N)})](1-x_1),    
\end{align}
where
\begin{align*}
\overline{\mathcal{A}}_j^{(N)} =&\ \frac{1}{\Pi_{i=1}^N\mathcal{Z}_i}\mathlarger{\sum_{n_{tot}=0}^\infty}\ ^{(N+n_{tot}-1)}C_{n_{tot}} \times 2n_{tot}\Delta^2 \Bigg(\frac{1}{\eta_j^{(N)}}\Bigg)^2\nonumber\\
&\hspace{1.5cm}\times\exp{(-\gamma_j\, n_{tot}\hbar\omega_j})\nonumber\\
=&\ (1-e^{-\gamma_j\hbar\omega_j})^N z_j^{-\frac{\beta_j ^2}{4 \Delta ^2}}~\frac{N}{2} B_{z_j}\Bigg(\frac{\beta_j ^2}{4 \Delta ^2}+1,-N\Bigg)
\end{align*}
and 
\begin{align*}
\overline{\mathcal B}_j^{(N)}=&\ \frac{1}{\Pi_{i=1}^N\mathcal{Z}_i}\mathlarger{\sum_{n_{tot}=0}^\infty}\ ^{(N+n_{tot}-1)}C_{n_{tot}} \times 2(n_{tot}+1)\Delta^2\Bigg(\frac{1}{\eta_j'^{(N)}}\Bigg)^2\nonumber\\
&\hspace{1.5cm}\times\exp{(-\gamma_j\, n_{tot}\hbar\omega_j})\nonumber\\
=&\ (1-e^{-\gamma_j\hbar\omega_j})^N z_j^{-\frac{\beta ^2}{4 \Delta ^2}-N}\frac{1}{2}\Bigg[B_{z_j}\Bigg(\frac{\beta_j ^2}{4 \Delta ^2}+N,1-N\Bigg)\nonumber\\
&\hspace{1.5cm}+N B_{z_j}\Bigg(\frac{\beta_j ^2}{4 \Delta ^2}+N+1,-N\Bigg)\Bigg],
\end{align*}
with $j=(c,h)$, $z_j=e^{-\hbar\gamma_j\omega_j}$, and $B_z(a, b) = \int_0^z t^{a-1} (1-t)^{b-1} \, dt$ is the incomplete Beta function.

% \begin{equation}
% \begin{array}{ll}
%     \frac{\overline{\mathcal{A}}_h^{(N)}(1-\tau)-[\overline{\mathcal{A}}_h^{(N)}(1-\chi_c^{(N)})+\overline{\mathcal{A}}_c^{(N)}(1-\overline{\mathcal{A}}_h^{(N)})]}{\chi_h^{(N)}(1-\tau)-[\chi_h^{(N)}(1-\overline{\mathcal{A}}_c^{(N)})+\chi_c^{(N)}(1-\chi_h^{(N)})]}\leq \frac{x_1}{1-x_1},\\
%   \\
%     \frac{\chi_h^{(N)}(1-\overline{\mathcal{A}}_c^{(N)})+\chi_c^{(N)}(1-\chi_h^{(N)})}{\overline{\mathcal{A}}_h^{(N)}(1-\chi_c^{(N)})+\overline{\mathcal{A}}_c^{(N)}(1-\overline{\mathcal{A}}_h^{(N)})}\geq \frac{x_1}{1-x_1}.
% \end{array}
% \end{equation}

\begin{figure}[!b]
    \centering
    \includegraphics[width=\columnwidth]{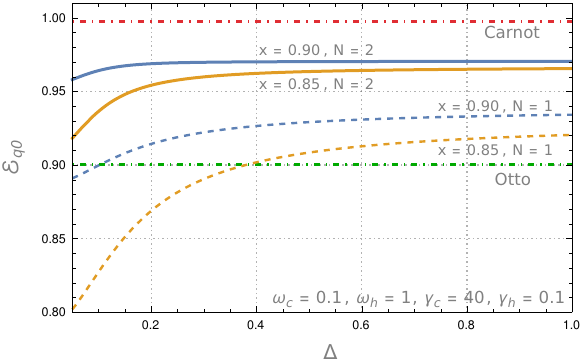}
    \caption{\label{fig:eff1}The efficiency of a single quasi-Otto cycle for $N=1$ and $N=2$ bath modes and two different initial states with $x_1=0.9$ (blue) and $x_1=0.85$ (yellow) with respect to a varying interaction strength, $\Delta$. It demonstrates that, for judicious choice of the parameters and the initial states, the quasi-cycle can be more efficient than the Otto cycle (green horizontal line) because of the non-Markovian, memory-dependent, and non-equilibrium nature of its dynamics. The efficiency can, however, never break the Carnot bound (red horizontal line). We use the natural unit system and set $\hbar=k_B=1$.}
%    , but lower than the Carnot cycle (red horizontal line). The figure shows that, unlike Carnot and ideal Otto cycle, where exact thermalisation was considered to be achievable, the choice of input state changes the efficieny of the cycle. This interesting phenomena is the consequence of non-Markovian, memory dependent and non-equilibrating nature of the dynamics. Furthermore, we see that although the quasi-cycle always obeys the Carnot bound, it can generally achieve a better efficiency than its ideal counterpart. 
    
\end{figure}
\begin{figure}[!t]
    \centering
    \includegraphics[width=\columnwidth]{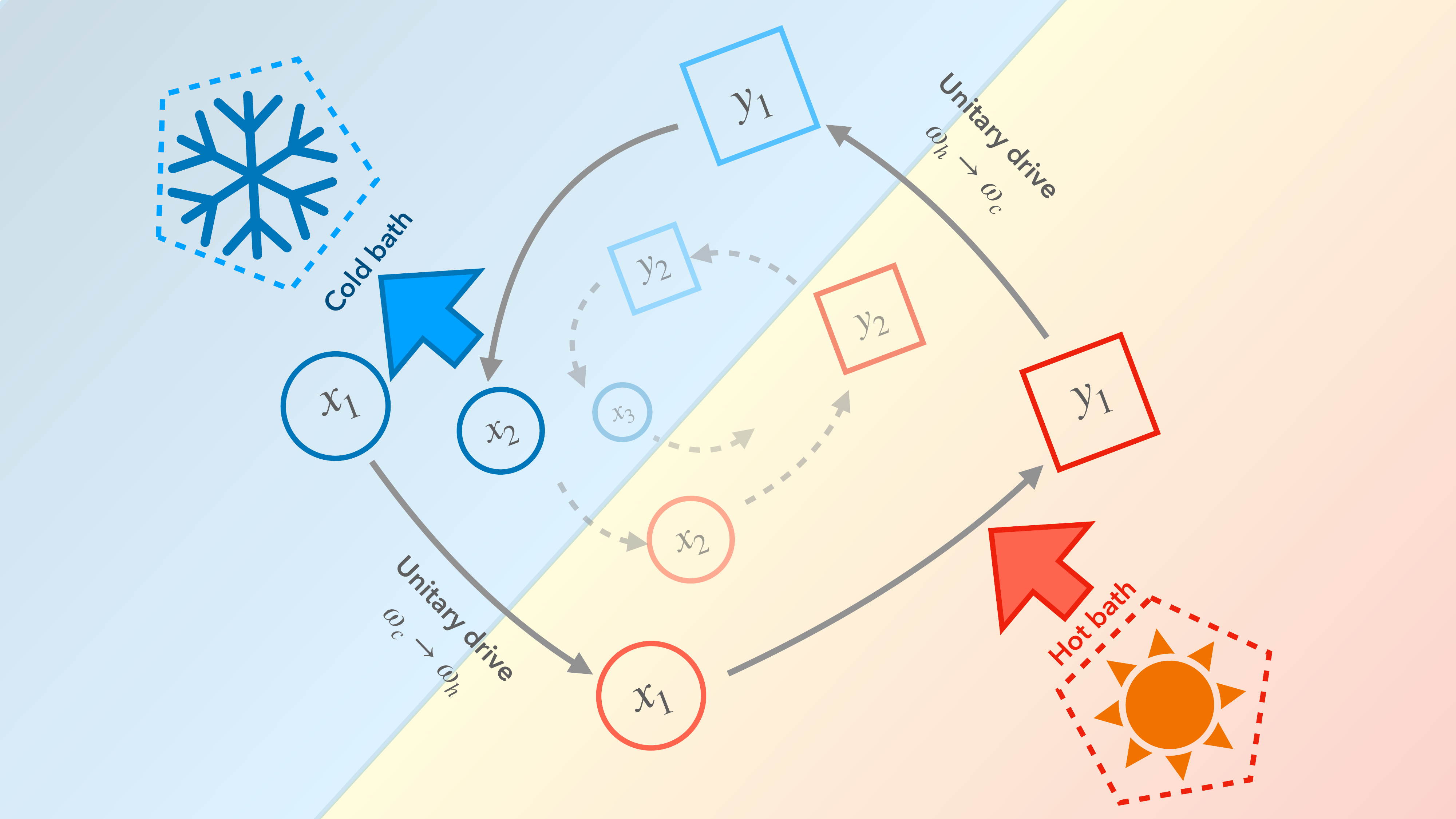}
    \caption{Schematic diagram of the proposed multi-cyclic quasi-Otto engine.}
    \label{fig:multirunschmatic}
\end{figure}
\begin{figure}[!t]
    \centering
    \includegraphics[width=\columnwidth]{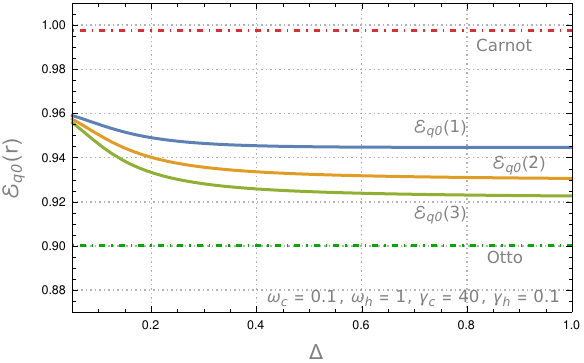}
    \caption{$N=1$, $x_1=0.95$ In this plot, we demonstrate variation of The efficiency of the quasi-Otto cycle after $i$-th cycle.with interaction strength $\Delta$. We have considered the efficiency of one, two and three cycles denoted by $\zeta_1,~\zeta_2,~\zeta_3$ respectively. We have taken $N=1$, $x_1=0.95$. We see that with the increasing number of cycles, the efficiency is decreasing, but remains greater than the Otto efficiency. Though it is also tending to saturate with the increasing number of cycles. We us the natural unit system and set $\hbar=k_B=1$.}
    \label{fig:eff2}
\end{figure}

For the specific case when the Otto cycle converges to the Carnot cycle, the condition for the heat engine regime can be expressed as
\begin{equation}
    \frac{\overline{\mathcal B}_h^{(N)}(1-\overline{\mathcal{A}}_c^{(N)})+\overline{\mathcal B}_c^{(N)}(1-\overline{\mathcal B}_h^{(N)})}{\overline{\mathcal{A}}_h^{(N)}(1-\overline{\mathcal B}_c^{(N)})+\overline{\mathcal{A}}_c^{(N)}(1-\overline{\mathcal{A}}_h^{(N)})}\leq \frac{x_1}{1-x_1},
\end{equation}
since, in this case, $x_2\leq x_1$ [i.e., after completion of the first (quasi-)cycle, the final state cannot be less energetic than the initial state where it started]. %When $x_2=x_1$, the cycle is perfect and the efficiency will reach Carnot bound. 
We are interested in the case when the Otto cycle is strictly less efficient than Carnot, i.e., $\gamma_h\omega_h<\gamma_c\omega_c$, for then, the quasi-Otto cycle can be more efficient than its ideal counterpart (while obeying the Carnot bound) when the following two conditions are satisfied:
\begin{align}
 \frac{x_1}{1-x_1} \geq&\ \frac{\overline{\mathcal{A}}_h^{(N)}(1-\tau)-[\overline{\mathcal{A}}_h^{(N)}(1-\overline{\mathcal B}_c^{(N)})+\overline{\mathcal{A}}_c^{(N)}(1-\overline{\mathcal{A}}_h^{(N)})]}{\overline{\mathcal B}_h^{(N)}(1-\tau)-[\overline{\mathcal B}_h^{(N)}(1-\overline{\mathcal{A}}_c^{(N)})+\overline{\mathcal B}_c^{(N)}(1-\overline{\mathcal B}_h^{(N)})]},\nonumber\\ 
 \frac{x_1}{1-x_1}\leq&\ \frac{\overline{\mathcal B}_h^{(N)}(1-\overline{\mathcal{A}}_c^{(N)})+\overline{\mathcal B}_c^{(N)}(1-\overline{\mathcal B}_h^{(N)})}{\overline{\mathcal{A}}_h^{(N)}(1-\overline{\mathcal B}_c^{(N)})+\overline{\mathcal{A}}_c^{(N)}(1-\overline{\mathcal{A}}_h^{(N)})}.
\end{align}
with $\tau=\gamma_h\omega_h/\gamma_c\omega_c$. In Fig.~\ref{fig:eff1}, we show how choosing a suitable initial state lets the quasi-Otto engine obtain a greater efficiency than the Otto engine.  

\subsection{Efficiency of the quasi-Otto engine: Multiple runs}
\noindent
Since, unlike the Otto cycle, the state at the end of a quasi-cycle is not the same as the one it started with (e.g., $x_2\neq x_1$), it is interesting to consider the efficiency after multiple runs of the quasi-Otto cycle. For more than one cycle, the final state of the previous cycle acts as the input for the next cycle. We show the schematic of the multi-running quasi-cycle in Fig.~\ref{fig:multirunschmatic}.  It is simple to extrapolate Eq.~\eqref{eff1} and calculate the efficiency after $r$ runs of the quasi-cycle:
\begin{equation}
    \mathcal{E}_{qO}(r) = 1-\frac{\omega_c}{\omega_h}\left|\frac{\sum_{i=1}^r(y_i-x_{i+1})}{\sum_{i=1}^r(y_i-x_i)}\right|. \label{multeff1}
\end{equation}
If we had a perfect cycle, i.e., $x_{i+1}=x_i$, then, like the ideal Otto engine, the efficiency would not have changed with $r$. In our case, the non-equilibrium nature of the dynamics hinders the system from returning to the initial state. As we can see from Fig.~\ref{fig:eff2}, the efficiency decreases with increasing $r$, but the rate of decrease reduces with $r$. However, it is interesting to note that asymptotically, in the limit $r\to \infty$, the efficiency of the quasi-cycle tends to that of the quantum Otto cycle:
\begin{equation}
    \lim_{r\to\infty} \mathcal E_{qO}(r) = \mathcal E_O.
\end{equation}
Since $x_i$'s and $y_i$'s are fractions and $x_{i+1}$ and $x_i$ come more and more closer to each other as $i$ increases, $\lim_{r\to\infty}\left|{\sum_{i=1}^r(y_i-x_{i+1})}/{\sum_{i=1}^r(y_i-x_i)}\right|\to 1$. This implies that the quasi-cycle will be as efficient as the Otto engine if we run it many, many times.

% rotation and would have also been equal to $\zeta_0$. But, because of this non-equilibrium phenomena which does not allow the target state of each individual cycle rotate back to the input state, causes the variation of efficiency with more and more rotation. From Fig. \ref{fig:eff2}, it is evident that with increasing number of cycles the efficiency of the engine diminishes, although it also tend to saturate after a few rotation. Even then, with a suitable choice of input state, the engine remains more efficient than the ideal Otto engine. It is therefore apparent the non-equilibrating nature of the interaction is playing an important role in achieving a better efficient heat engine. In Fig. \ref{fig:multirunschmatic}, we have pictorially narrated the multi-cyclic rotation of the quasi-Otto engine. By comparing with Fig. \ref{fig:ottoPic}, we can clearly visualise the difference between the ideal and the quasi Otto cycles.

\section{Summary and conclusions}\label{sec5}
\noindent
In this paper, we constructed a dynamical model for a qubit system inside an optical cavity. The qubit interacts with vibrational modes of the cavity via the Jaynes-Cummings Hamiltonian. Without relying on the Born-Markov approximation, we derived the exact dynamics of this system in terms of the reduced map and obtained the corresponding Lindblad-type master equation, giving us a non-Markovian analogue of quantum dynamics based on the spin-boson model. Our analysis showed how information backflow from bath to system could dominate its evolution, keeping it from equilibrating with the environment. Due to the memory-dependent nature of the dynamics, the output quantum state depends on the choice of the input state, even after a long period.

This nature of the dynamics played an important role in a cavity-based quantum Otto cycle-like process we developed. Since the cycle is not closed like the Otto cycle due to the non-equilibrating nature of the dynamics, we call it the `quasi-Otto cycle'. We showed how, with some choices of the input state and parameters, it can be more efficient as a quantum engine than the Otto cycle. Interestingly, as the quasi-cycle is repeated multiple times, the efficiency of the quasi-Otto engine asymptotically approaches that of the Otto engine.

Our analysis indicates that non-equilibrium quantum systems can show thermodynamic advantages in quantum devices and, hence, play an important role in state-of-the-art quantum device engineering. Our findings also encourage further exploration in quantum electrodynamics-based theory of opto-mechanical systems in conjunction with modelling experimental protocols. Jaynes-Cummings-type interactions exclude the strong coupling regime, resulting in energy conservation in the photon-phonon dynamics under resonance. The quasi-Otto engine hints that investigating the strong coupling regime could lead us to even more interesting non-conservative interactions exhibiting highly non-Markovian, non-equilibrium phenomena with significant effects on quantum heat devices. We plan to investigate such effects in quantum device engineering in future.

\section{Acknowledgements}
\noindent 
SM and SB acknowledge support from the Ministry of Electronics and Information Technology (MeitY), Government of India, under Grant No. 4(3)/2024-ITEA.

\begin{appendix}\label{app}
    
\section{}\label{appA}
\noindent
In this section, we elaborate on the case of a qubit interacting with a multimode cavity. For clarity, we describe the dynamics of the qubit interacting with two modes. By interpolation, one can then reach the case of $N$-mode interactions described in Sec.~\ref{NmodeModel}.
% \subsection{A qubit interacting with two bath modes}
% \noindent
Let us consider a qubit interacting with two bosonic modes with the same frequency. The total Hamiltonian for this setup can be written as
\begin{align}
 H_T^{(2)} =&\ \hbar\omega_0\sigma_z + \hbar\omega a_1^\dagger a_1+\hbar\omega a_2^\dagger a_2 + \hbar\Delta \left(\sigma_-a_1^\dagger+\sigma_+a_1\right)\nonumber\\
&\ +\hbar\Delta \left(\sigma_-a_2^\dagger+\sigma_+ a_2\right),   \label{8}
\end{align}
where we assume that both modes interact with the same interaction strength, $\Delta$. We want to solve the corresponding Schr\"{o}dinger equation with the following restriction on the allowed transitions:
\begin{align}
\ket{0}\ket{n_1}\ket{n_2}\rightarrow&\  A_t^{(2)}\ket{0}\ket{n_1}\ket{n_2}+B_{1,t}^{(2)}\ket{1}\ket{n_1-1}\ket{n_2}\nonumber \\
&\ +B_{2,t}^{(2)}\ket{1}\ket{n_1}\ket{n_2-1}=\ket{\psi^{(2)}},\nonumber\\
\ket{1}\ket{n_1}\ket{n_2}\rightarrow&\  C_t^{(2)}\ket{1}\ket{n_1}\ket{n_2}+D_{1,t}^{(2)}\ket{0}\ket{n_1+1}\ket{n_2}\nonumber\\
&\ +D_{2,t}^{(2)}\ket{0}\ket{n_1}\ket{n_2+1}=\ket{\chi^{(2)}}.  \label{9}  
\end{align}
As mentioned in Sec.~\ref{NmodeModel}, here, we have assumed the photonic modes do not interact with each other directly (a reasonable assumption since we do not consider photons that are highly energetic and can polarise the vacuum); they only interact via the system (otherwise, the above closed-form transition may not hold in general). 
%Before proceeding further, it is important to note that upon acting the Hamiltonian $H_T^{(2)} $ on the states $(\ket{\psi^{(2)}},\ket{\chi^{(2)}})$, we get a series of unwanted extra states representing various transitions between the bath states $\ket{n_1}$ and $\ket{n_2}$, rendering $(\ket{\psi^{(2)}},\ket{\chi^{(2)}})$to be improper solutions of the corresponding Schr\"{o}dinger equation associated to the Hamiltonian given in \eqref{8}. These unwanted transitions are caused by the interaction Hamiltonian between the qubit and the two modes. To encounter this difficulty, we consider a further restriction on the dynamics that the bath modes are highly energetic with sufficiently large characteristic frequencies in comparison to the interaction frequency. Hence, the bath modes evolve so fast compared to the interaction time period that it does not allow any transition between the bath modes. This restriction on the allowed transitions enables us to get the following sets of coupled differential equations from the system-bath dynamics. 

For the state $\ket{\psi^{(2)}}$, we get the following set of coupled differential equations:
\begin{align}
\dot{A}_t^{(2)} =&\ -i\left(\omega_0+(n_1+n_2)\omega\right)~A_t^{(2)} -i\Delta\sqrt{n_1}~B_{1,t}^{(2)} -i\Delta\sqrt{n_2}~B_{2,t}^{(2)},~~\nonumber\\
\dot{B}_{1,t}^{(2)} =&\ -i\left(-\omega_0+(n_1+n_2-1)\omega\right)~B_{1,t}^{(2)} -i\Delta\sqrt{n_1}~A_t^{(2)},\nonumber\\
\dot{B}_{2,t}^{(2)} =&\ -i\left(-\omega_0+(n_1+n_2-1)\omega\right)~B_{2,t}^{(2)} -i\Delta\sqrt{n_2}~A_t^{(2)}.  \label{10}  
\end{align}
Solving these, we get
\begin{align}
A_t^{(2)} =&\ e^{-i\alpha^{(2)} t/2}\left(\cos{\left(\eta^{(2)} t/2\right)}-i\frac{\beta^{(2)}}{\eta^{(2)}}\sin{\left(\eta^{(2)} t/2\right)}\right)\nonumber\\
B_{1,t}^{(2)} =&\ -2i\Delta\sqrt{n_1}~e^{-i\alpha^{(2)} t/2}\left(\frac{\sin{\left(\eta^{(2)} t/2\right)}}{\eta^{(2)}}\right),\nonumber\\
B_{2,t}^{(2)} =&\ -2i\Delta\sqrt{n_2}~e^{-i\alpha^{(2)} t/2}\left(\frac{\sin{\left(\eta^{(2)} t/2\right)}}{\eta^{(2)}}\right), \label{11}   
\end{align}
with
\begin{align*}
\alpha^{(2)} =&\ (2(n_1+n_2)-1)\omega,\quad \beta^{(2)}= 2\omega_0+\omega,\\
\eta^{(2)} =&\ \sqrt{(\beta^{(2)})^2+4(n_1+n_2)\Delta^2}.    
\end{align*}
One can easily verify that $|A_t^{(2)}|^2+|B_{1,t}^{(2)}|^2+|B_{2,t}^{(2)}|^2=1$. Similarly, we can solve for the state $\ket{\chi^{(2)}}$ to get,
\begin{align}
C_t^{(2)} =&\ e^{-i\alpha'^{(2)} t/2}\left(\cos{\left(\eta'^{(2)} t/2\right)}-i\frac{\beta'^{(2)}}{\eta'^{(2)}}\sin{\left(\eta'^{(2)} t/2\right)}\right)\nonumber\\
D_{1,t}^{(2)} =&\ -2i\Delta\sqrt{n_1+1}~e^{-i\alpha'^{(2)} t/2}\left(\frac{\sin{\left(\eta'^{(2)} t/2\right)}}{\eta'^{(2)}}\right),\nonumber\\
D_{2,t}^{(2)} =&\ -2i\Delta\sqrt{n_2+1}~e^{-i\alpha'^{(2)} t/2}\left(\frac{\sin{\left(\eta'^{(2)} t/2\right)}}{\eta'^{(2)}}\right),    
\end{align}
with
\begin{align*}
\alpha'^{(2)} =&\ (2(n_1+n_2)+1)\omega,\quad \beta'^{(2)}= -(2\omega_0+\omega),\\
\eta'^{(2)} =&\ \sqrt{(\beta'^{(2)})^2+4(n_1+n_2+2)\Delta^2}.    
\end{align*}
and $|C_t^{(2)}|^2+|D_{1,t}^{(2)}|^2+|D_{2,t}^{(2)}|^2=1$. As we did in the single-mode case, we can now derive the dynamical map for the qubit as
\begin{align}
\rho_{00}(t) =&\ (1-\mathcal{A}_t^{(2)})~\rho_{00}+\mathcal{B}_t^{(2)}~\rho_{11},\nonumber\\
\rho_{11}(t) =&\ (1-\mathcal{B}_t^{(2)})~\rho_{11}+\mathcal{A}_t^{(2)}~\rho_{00},\nonumber\\
\rho_{01}(t) =&\ \mathcal{C}_t^{(2)}~\rho_{01},~~\rho_{10}(t)=\rho_{01}^*(t), \label{13}   
\end{align}
with 
\begin{align}
\mathcal{A}_t^{(2)} =&\ \frac{1}{\mathcal{Z}_1\mathcal{Z}_2}\sum_{n_i=0}^\infty 4(n_1+n_2)\Delta^2 \left(\frac{\sin{\left(\eta^{(2)} t/2\right)}}{\eta^{(2)}}\right)^2\nonumber\\
&\ \quad\times\exp{(-\gamma (n_1+n_2)\hbar\omega}),\nonumber\\ 
\mathcal{B}_t^{(2)} =&\ \frac{1}{\mathcal{Z}_1\mathcal{Z}_2}\sum_{n=0}^\infty 4(n_1+n_2+1)\Delta^2\left(\frac{\sin{\left(\eta'^{(2)} t/2\right)}}{\eta'^{(2)}}\right)^2\nonumber\\
&\ \quad\times\exp{(-\gamma (n_1+n_2)\hbar\omega}),\nonumber\\
\mathcal{C}_t^{(2)} =& \frac{e^{-i\omega t}}{\mathcal{Z}_1\mathcal{Z}_2}\sum_{n=0}^\infty \left(\cos{\left(\eta^{(2)} t/2\right)}-i\frac{\beta^{(2)}}{\eta^{(2)}}\sin{\left(\eta^{(2)} t/2\right)}\right)\nonumber\\
&\ \quad\times\left(\cos{\left(\eta'^{(2)} t/2\right)}+i\frac{\beta'^{(2)}}{\eta'^{(2)}}\sin{\left(\eta'^{(2)} t/2\right)}\right)\nonumber\\
&\ \quad\times\exp{(-\gamma (n_1+n_2)\hbar\omega}).
\end{align}

\section{Constructing the Lindblad-type master equation}\label{appB}
\noindent
Here we construct the Lindblad-type master equation for the evolution of a qubit interacting with an arbitrary but finite number of photonic modes. Given an invertible completely positive trace-preserving linear dynamical map, it is always possible to construct a master equation whose solution is the map itself~\citep{master1,*master2}. Since the map in Eq.~\eqref{17} is invertible, we can construct a Lindblad-type master equation for it. Let us express the dynamical map as
\begin{equation}\label{dynamicalmap}
    \rho (t) = \Omega [\rho (0)].
\end{equation}
It corresponds to a Lindblad-like master equation of the form
\begin{equation}\label{eqom}
    \Dot{\rho} (t) = \widetilde{\Lambda} [\rho (t)],
\end{equation}
with $\widetilde{\Lambda} [.]$ to be the generator of the given map. %Invertibility of an operation ensures that if the transformation $\sigma = \Omega(\rho)$ is given, then the inverse transformation $\rho = \Omega^{-1}(\sigma)$ exists for all $\rho$.
Now, we adopt the method of Ref.~\citep{master1,master2} and consider $\lbrace \mathcal{G}_i\rbrace$ to be a given Harmitian orthonormal basis with $\mathcal{G}_0 = \mathbb{I}/\sqrt{2}$ and $\mathcal{G}_i^{\dagger} = \mathcal{G}_i$. All $\mathcal{G}_i$'s except $\mathcal{G}_0$ are traceless and satisfy $\text{Tr}[\mathcal{G}_i\mathcal{G}_j] = \delta_{ij}$. In this basis, we can express the operation in Eq.~\eqref{dynamicalmap} as
\begin{equation*}
    \Omega [\rho (0)] = \sum_{m,\,n} \text{Tr}[\mathcal{G}_m\cdot \Omega [\mathcal{G}_n]]~ \text{Tr}[\mathcal{G}_n\cdot \rho (0)]~\mathcal{G}_m = [F(t)s(0)]~\mathcal{G}^T,
\end{equation*}
where $F_{mn}=\text{Tr}[\mathcal{G}_m\cdot \Omega [\mathcal{G}_n]]$ and $s_n(t) = \text{Tr}[\mathcal{G}_n\cdot \rho(t)]$. Calculating the derivative with respect to time, we get
\begin{equation*}
    \dot{\rho}(t) = [\dot{F}(t)s(0)]~ \mathcal{G}^T.
\end{equation*}
Let us consider a matrix $L$ with elements $L_{mn} = \text{Tr}[\mathcal{G}_m\cdot \widetilde{\Lambda} [\mathcal{G}_n]]$. Hence, we can now express Eq.~\eqref{eqom} as
\begin{equation}\label{Eq13}
    \dot{\rho}(t) = \sum_{m,n} \text{Tr}[\mathcal{G}_m]\widetilde{\Lambda} [\mathcal{G}_n] \text{Tr}[\mathcal{G}_n\cdot\rho (t)]\ \mathcal{G}_m = [L(t)s(t)]\ \mathcal{G}^T.
\end{equation}
It is possible to verify that
\begin{equation*}
    \dot{F}(t) = L(t) F(t) \implies L(t)= \dot{F}(t) F(t)^{-1}.
\end{equation*}
We observe that to obtain $L(t)$, $F(t)^{-1}$ must exist and $F(0) = \mathbb{I}$. Once both invertibility and the initial conditions are satisfied, we obtain the $L$ matrix and the master equation in the form of Eq.~\eqref{Eq13} follows directly from it. The master equation can be written in the Lindblad form~\citep{master1,master2}.

Using the above protocol, we construct the exact master equation for the dynamics of a two-level system interacting with $N$ bosonic modes as,
\begin{align}
\frac{d}{dt}\rho(t) =&\ i [\rho(t),~U(t)\sigma_z]+\Gamma_{dep}(t)\left( \sigma_z\rho(t)\sigma_z-\rho(t) \right) \nonumber\\
&\ +\Gamma_{d}(t)\left(\sigma_-\rho(t)\sigma_+-\frac{1}{2}\{\sigma_+\sigma_-,~\rho(t)\}\right)\nonumber\\
&\ +~\Gamma_{a}(t)\left(\sigma_+\rho(t)\sigma_--\frac{1}{2}\{\sigma_-\sigma_+,~\rho(t)\}\right)
\end{align}
with
\begin{align*}
U(t) =&\ \frac{1}{4}\frac{(\mathcal{C}_t^{(N)})_I}{(\mathcal{C}_t^{(N)})_R}\frac{d}{dt}\Bigg[\ln{\Big(1+\bigg(\frac{(\mathcal{C}_t^{(N)})_R}{(\mathcal{C}_t^{(N)})_I}\bigg)^2\Big)}\Bigg],\nonumber\\  
\Gamma_{dep}(t) =&\ \frac{1}{4}\frac{d}{dt}\left[\ln\left(\frac{1-\mathcal{A}_t^{(N)}-\mathcal{B}_t^{(N)}}{|\mathcal{C}_t^{(N)}|^2}\right)\right],\\
\Gamma_{d}(t) =&\ \Bigg[\frac{d}{dt}\left(\frac{\mathcal{A}_t^{(N)}-\mathcal{B}_t^{(N)}}{2}\right)\\
&\ -\left(\frac{\mathcal{A}_t^{(N)}-\mathcal{B}_t^{(N)}+1}{2}\right)\frac{d}{dt}\ln(1-\mathcal{A}_t^{(N)}-\mathcal{B}_t^{(N)})\Bigg],\\
\Gamma_{a}(t) =&\ -\Bigg[\frac{d}{dt}\left(\frac{\mathcal{A}_t^{(N)}-\mathcal{B}_t^{(N)}}{2}\right)\\
&\ -\left(\frac{\mathcal{A}_t^{(N)}-\mathcal{B}_t^{(N)}-1}{2}\right)\frac{d}{dt}\ln(1-\mathcal{A}_t^{(N)}-\mathcal{B}_t^{(N)})\Bigg],
\end{align*}
where $R$ and $I$ in the subscripts indicate the real and imaginary parts of the function, respectively. The above Lindblad-type master equation is constructed without the Born-Markov or stationary bath approximation~\citep{book1}. Hence, it is an example of an exact master equation for a non-Markovian spin-boson model of system-bath coupling with direct applicability in experiments in spin-cavity interactions.  

\end{appendix}

% \bibliography{apssamp}% Produces the bibliography via BibTeX.
%\bibliographystyle{apsrev4-1}
\bibliography{reference}

\end{document}